\documentclass[aps,pra,twocolumn]{revtex4-2}
\usepackage{amsfonts}
\usepackage{amsmath}
\usepackage{mathrsfs}
\usepackage{amssymb}
\usepackage{graphicx}
\usepackage{bm}
\usepackage{color}
\usepackage[colorlinks=true,linkcolor=blue,anchorcolor=blue,citecolor=blue,urlcolor=blue]{hyperref}

\begin{document}
\preprint{APS/123-QED}

\title{Superfluidity breakdown of Rabi-coupled two-component Bose-Einstein condensates in optical lattices}

\author{Huaxin He}
\affiliation{International Center of Quantum Artificial Intelligence for Science and Technology (QuArtist) and Department of Physics, Shanghai University, Shanghai 200444, China}

\author{Yongping Zhang}
\email{yongping11@t.shu.edu.cn}
\affiliation{International Center of Quantum Artificial Intelligence for Science and Technology (QuArtist) and Department of Physics, Shanghai University, Shanghai 200444, China}

\begin{abstract}
	
We demarcate the unstable regimes of superfluids in a Rabi-coupled two-component Bose-Einstein condensate in the presence of optical lattices. We find that the Rabi coupling can stabilize superfluids. A significant separation between Landau and dynamical instabilities is presented in a Rabi-coupled Zeeman lattice.

\end{abstract}

\maketitle

\section{Introduction}

Atomic Bose-Einstein condensates (BECs) loaded into optical lattices~\cite{Morsch2006} offer versatile and important quantum many-body platforms~\cite{Bloch2008}.  The study on lattice BECs not only enables good services for purposes of quantum simulation~\cite{Georgescu2014} but also provides fundamental physics which are  essentially relevant to the interplay between quantum many-body interactions and periodic potentials. Among these physics, lattice superfluidity is of interest~\cite{Wu2001, Burger2001,Smerzi2002,Konotop2002, Machholm2003, Wu2003,Menotti2003,Fallani2004,Modugno2004,Sarlo2005, Danshita2007, Hui2011, Chen2011,Shaoliang2013,Xu2013, Dasgupta2016}.  It has been predicted  that there are two different mechanisms for the breakdown of lattice superfluidity~\cite{Wu2001,Wu2003}. They are dynamical and Landau instabilities, both of which are relevant to collective excitations~\cite{Wu2001,Fallani2004,Modugno2004,Chen2010}.
In the presence of optical lattices, the systems feature Bloch band gap structures associated with Bloch states. The precise controllability of optical lattices makes it possible to experimentally load BECs into selected Bloch states in an arbitrary band~\cite{Denschlag2002}.
With many-body interactions, the Bloch state can be characterized by a set of collective excitations~\cite{Ozeri2005,Fabbri2009}. If there are some modes in collective excitations growing up exponentially, the corresponding Bloch state is dynamically unstable. While, Landau instability occurs if some modes in collective excitations possess negative energies, in this situation, the relevant Bloch state is not the local minimum of energy functional and is energetically unstable. Both the instabilities for destroying lattice superfluids have been experimentally confirmed and explored~\cite{Fallani2004,Sarlo2005}.

The study of lattice superfluidity has been generalized from a single-component BEC to multiple components~\cite{Jin2005,Hooley2007,Liu2007,Ruostekoski2007, Barontini2009, Huang2010,Watanabe2018}. Multi-component BECs possess more degrees of freedom comparing with the single component. The presence of inter-component interactions gives instabilities rich structures~\cite{Barontini2009} and can stabilize certain Bloch states that are unstable in single-component analogues~\cite{Ruostekoski2007,Watanabe2018}.

In the present paper, we study the breakdown of lattice superfluidity in the two-component BEC that is linearly and coherently Rabi-coupled. Ever since the achievement of atomic BECs, the Rabi coupling has been an outstandingly experimental means to coherently control over population between components and to introduce new phenomena, such as bringing topological defects into BECs~\cite{Wright2009}.  Depending on the energy splittings between components and experimental purposes, the Rabi coupling  can be experimentally implemented  via the interactions between matters and external fields, including microwave~\cite{Hamner2013}, radio-frequency radiations~\cite{Lundblad2008} and lights~\cite{Matthew1998}. Collisional interactions of two-component BECs do not exchange particles between components, while the Rabi coupling prefers population balance via components mixing. There is a competition between interactions and the Rabi coupling. The Rabi-coupled two-component BECs represent active platforms for exploring nonlinear phenomena~\cite{Lee2009,Adhikari2009,Sabbatini2011,Zhan2014,Usui2015,Sartori2015,Congy2016,Abad2013,Qu2017,Bornheimer2017,Uranga2018,Baals2018,Ihara2019,Eto2020, Farolfi2020,Fan2020,Momme2020}. In recent experiments, the Rabi coupling is accompanied by the generation of artificial spin-orbit coupling~\cite{Lin2011}. A spin-orbit-coupled and Rabi-coupled two-component BEC has been loaded into an optical lattice~\cite{Hamner2015}. Dynamical instability of this joint system has experimentally and theoretically  shown the asymmetry of superfluidity originating from  spin-orbit coupling~\cite{Hamner2015, Zhang2013,Toniolo2014}.

The breakdown of lattice superfluidity in a Rabi-coupled BEC is analyzed from Landau and dynamical instabilities. Three typical  configurations of optical lattices are studied, they are spin-independent lattice, Zeeman lattice and matter grating. The interplay between interactions, the Rabi coupling and optical lattices generates interesting features in the instabilities of lattice superfluids. We find that the Rabi coupling can play an important role in stabilizing superfluids which are unstable in the absence of the Rabi coupling. There is an obvious competition between
the Zeeman lattice and the Rabi coupling. The coexistence of them results in a particular translational symmetry, which changes the spectrum in a very interesting way. Due to this symmetry, the regime of the Landau instability is significantly separated from the dynamical instability. This paper is organized as follows. In Sec.~\ref{model}, we present the theoretical frame for the stability analysis of nonlinear Bloch states. In Sec.~\ref{Independent}, we show the stabilization of superfluids by the Rabi coupling in the spin-independent optical lattice. The extension of first Brillouin zone and the significant separation between Landau and dynamical instabilities for the Rabi-coupled Zeeman lattice are demonstrated in Sec.~\ref{ZeemanLattice}. In Sec.~\ref{Mattergrating}, we discuss the instabilities of superfluids in the Rabi-coupled matter grating, and the conclusion follows in Sec.~\ref{Conclusion}.

\section{Model}
\label{model}

The dynamics of the Rabi-coupled two-component BEC in optical lattices is described by the following Gross-Pitaevskii equation (GPE),
\begin{align}
& i\frac{\partial \Psi_1}{\partial t}=\left[ H_1+ v_1 \cos(x) \right]  \Psi_1 +\frac{\Omega}{2}\Psi_2, \notag \\
& i\frac{\partial \Psi_2}{\partial t}=\left[ H_2+ v_2 \cos(x)  \right]\Psi_2+\frac{\Omega}{2} \Psi_1,
\label{GP}
\end{align}
with
\begin{align}
&H_1= -\frac{1}{2}\frac{\partial^2}{\partial x^2} +g|\Psi_1|^2 +g_{12}|\Psi_2|^2, \notag \\
&H_2= -\frac{1}{2}\frac{\partial^2}{\partial x^2} +g_{12}|\Psi_1|^2 +g|\Psi_2|^2.
\end{align}
The GPE  is quasi-one-dimensional on account of the tight traps in the transversal direction. $\Psi=(\Psi_1, \Psi_2)^T$ are the two-component wave functions.  The optical lattices are modeled as spin-dependent; the first component feels the depth of lattice as $v_1$ and the depth for the second component is $v_2$. The spin-dependent optical lattices can be experimentally realized in the hyperfine states  $|1,-1\rangle $ and $|2,-2\rangle $ of $^{87}$Rb atoms by dressing two far-detuned linearly polarized lasers~\cite{Mandel2003, Yang2017}. The incident angle between linear polarization of lasers plays an essential role, if the angle vanishes the lattices become spin-independent. Meanwhile, the two hyperfine states can be coupled by a microwave field~\cite{Mandel2003}. Such coupling causes the existence of the Rabi coupling in the above GPE. The strength of the Rabi coupling is $\Omega$.  The mean-field interactions are characterized by the coefficients $g_{ij}$ which are proportional to s-wave scattering lengths. For simplicity, we assume the coefficients of intra-component interactions are same and are equal to $g$. The coefficient of inter-component interactions is $g_{12}$. For the convenience of numerical calculations, the GPE is dimensionless. The unit of energy is $8E_{R}$ with $E_{R}=\hbar^2k_{L}^2/2m$, where $k_{L}$ is the wavevector of optical lattice lasers and $m$ is the mass of the $^{87}$Rb atom. The units of coordinate and time are chosen as $1/2k_{L}$ and $\hbar /8E_{R}$ respectively. In the presence of the Rabi coupling, the total atom number $\int_{0}^{2\pi} dx ( |\Psi_1|^2 +|\Psi_2|^2  )$ is conserved, here the integration is taken over a unit cell. We renormalize it as $1/(2\pi)\int_{0}^{2\pi} dx ( |\Psi_1|^2 +|\Psi_2|^2  )=1$.

The superfluidity of our system relates to the stability of Bloch states. We systematically study the instabilities of Bloch states to analyze superfluidity breakdown. The Bloch states are the stationary solutions of the GPE and are defined as
\begin{equation}
\Psi(x,t)=e^{-i\mu t+ikx}\begin{pmatrix} \Phi_1(x) \\  \Phi_2(x) \end{pmatrix},
\end{equation}
here, $\mu$ is the chemical potential and $k$ is the quasimomentum with the unit being $2k_L$. Periodic functions $\Phi(x)=(\Phi_1(x), \Phi_2(x))^T$ have the same period as optical lattices, $ \Phi(x)=\Phi(x+2\pi) $. It is worth noting that the two components have a same quasimomentum $k$ in the presence of the Rabi coupling.


The stability analysis of the Bloch state is performed by studying their collective excitations. To proceed, we add
perturbations into the stationary Bloch state,
\begin{align}
\label{Perturbation}
&\Psi_{1,2}=  \notag \\
&e^{-i\mu t+ikx}[\Phi_{1,2} (x)+U_{1,2}(x)e^{iqx-i\omega t} +V_{1,2}^*(x) e^{-iqx+i\omega^* t}], \notag
\end{align}
where $U_j(x)$ and $V_j(x)$ are perturbation amplitudes, $q$ and $\omega$ are the quasimomentum and energy of perturbations, respectively. The relationship $\omega(q)$ is collective excitation spectrum of the selected Bloch state $e^{-i\mu t+ikx}\Phi(x)$. After substituting the above wave functions $\Psi$ into the GPE and keeping only linear terms of perturbation amplitudes, we get the following Bogoliubov-de Gennes (BdG) equation,
\begin{equation}
\label{BdG}
\omega \begin{pmatrix} U_1 \\ V_1 \\U_2\\V_2\end{pmatrix}= \mathcal{H}_\text{BdG} \begin{pmatrix} U_1 \\ V_1 \\U_2\\V_2\end{pmatrix},
\end{equation}
with,
\begin{align}
  \mathcal{H}_\text{BdG}= \mathcal{L}+\frac{\Omega}{2} \begin{pmatrix} 0  & 0  &1 &0 \\ 0&0&0&-1\\ 1 &0&0&0\\ 0&-1&0&0\end{pmatrix},
\end{align}
and,
\begin{align}
&  \mathcal{L}= \notag \\
&
\left(\begin{array}{cccc}{  \mathcal{L}_{1}\left( k,q\right)}    &    {g\Phi_{1}^{2}}    &    {g_{12} \Phi_{2}^{*} \Phi_{1}}    &     {g_{12} \Phi_{1} \Phi_{2}}
\\   {-g \Phi_{1}^{*2}}           &        {-\mathcal{L}_{1}\left( -k,q\right)}      &      {-g_{12} \Phi_{2}^{*} \Phi_{1}^{*}}          &         {-g_{12} \Phi_{2} \Phi_{1}^{*}}
\\    {g_{12} \Phi_{1}^{*} \Phi_{2}}     &     {g_{12} \Phi_{1} \Phi_{2}}     &      {\mathcal{L}_{2}\left( k,q\right)}    &   {g \Phi_{2}^{2}}
\\    {-g_{12} \Phi_{1}^{*} \Phi_{2}^{*}}       &      {-g_{12} \Phi_{1} \Phi_{2}^{*}}      &     {-g \Phi_{2}^{*2}}    &       {-\mathcal{L}_{2}\left( -k,q\right)}\end{array}\right). \notag
\end{align}
Here, $\mathcal{L}_{1}( k,q)=-\frac{1}{2}[\frac{\partial}{\partial x}+i(k+q)]^2 +v_1\cos(x)-\mu+2g|\Phi_1|^2+g_{12}|\Phi_2|^2$  and  $\mathcal{L}_{2}( k,q)=-\frac{1}{2}[\frac{\partial}{\partial x}+i(k+q)]^2 +v_2\cos(x)-\mu+2g|\Phi_2|^2+g_{12}|\Phi_1|^2$. For a given Bloch state $\Phi$, we can calculate its collective excitation spectrum $\omega(q)$ by diagonalizing the BdG equation.  The Hamiltonian $\mathcal{H}_\text{BdG}$ in the BdG equation is non-Hermitian. Therefore, there is no guarantee that eigenvalues $\omega$ are real-valued. If some modes in $\omega$ become complex, the perturbation amplitudes grow up exponentially, which indicates that the corresponding Bloch state $\Phi$ is  dynamically unstable. Such dynamical instability eventually destroys the BEC Bloch states by blowing up perturbations exponentially.

The Landau instability relates to the collective excitation spectrum of $\tau_z  \mathcal{H}_\text{BdG}$~\cite{Wu2001,Machholm2003,Danshita2007,Shaoliang2013,Chen2010,Barontini2009}, where $\tau_z=\begin{pmatrix}
\sigma_z&0\\0& \sigma_z
\end{pmatrix} $ and $\sigma_z=\begin{pmatrix} 1&0  \\0 &-1\end{pmatrix} $.  $\tau_z  \mathcal{H}_\text{BdG}$ is Hermitian, so that, its excitation spectrum is real-valued. The collective excitation spectrum can be considered as the excited energy above a chosen Bloch state~\cite{Wu2001}. If some modes in the spectrum become negative, the chosen Bloch state is not energetically preferable. Such energetic instability is named as Landau instability.

For a homogeneous two-component BEC, the competition between intra- and inter-component interactions gives rise to phase separation instability: miscible states are unstable if  $\gamma >1$ with $\gamma= g_{12}/g$. In the presence of the Rabi coupling, phase separation occurs if $\gamma >1+\Omega/gn$, where $n$ is the density of the homogeneous BEC~\cite{Abad2013,Liu2007}.  The Rabi coupling dramatically changes the critical condition of the phase separation instability. In the following, we study the instabilities of Bloch states for three different configurations of optical lattices. In each configuration, we choose two typical sets of  interaction coefficients, $\gamma=3$ and $\gamma=1/3$, in order to emphasis on the role of the Rabi coupling.

\section{ Spin-independent lattice with $v_1=v_2$}
\label{Independent}

When $v_1=v_2$, the lattice is spin-independent. The instabilities of Bloch states in the spin-independent lattice without the Rabi coupling have been investigated~\cite{Jin2005,Liu2007,Ruostekoski2007}. In the absence of the Rabi coupling, the two components can carry different quasimomenta. For the same-quasimomentum-carrying Bloch states, the instability with miscible interactions $\gamma<1$ is analogous to that in the single-component case; all of them are dynamically unstable with immiscible interactions $\gamma>1$~\cite{Jin2005,Liu2007,Ruostekoski2007}. For the different-quasimomentum-carrying Bloch states, the instability exhibits rich features~\cite{Ruostekoski2007}.

\begin{figure}[b]
	\includegraphics[width=0.45\textwidth]{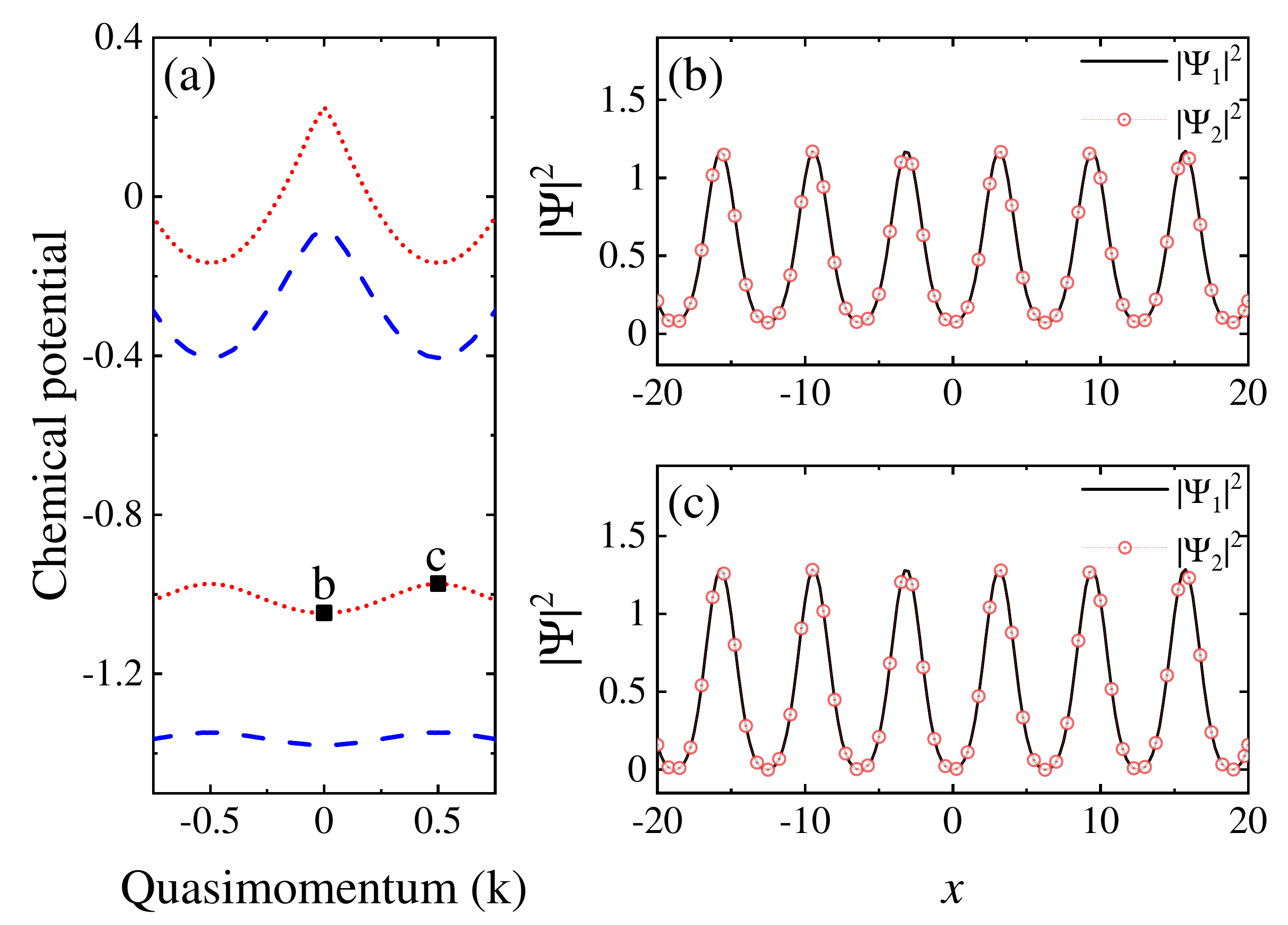}
	\caption{Nonlinear spectrum and nonlinear Bloch states in the Rabi-coupled spin-independent lattice, $v_1=v_2=0.5$ and $\Omega=1$. (a) Nonlinear spectrum (dotted red lines) and linear spectrum (dashed blue lines). Only the lowest two bands are shown. Nonlinear spectrum at $g=0.15,\gamma=g_{12}/g=1/3$ and at  $g=0.05,\gamma=3$ are exactly same. (b), (c) Nonlinear Bloch states in the lowest band at Brillouin zone center and edge respectively [labeled by squares in (a)].  }
	\label{Onespectrum}
\end{figure}

\begin{figure}[t]
	\includegraphics[width=0.45\textwidth]{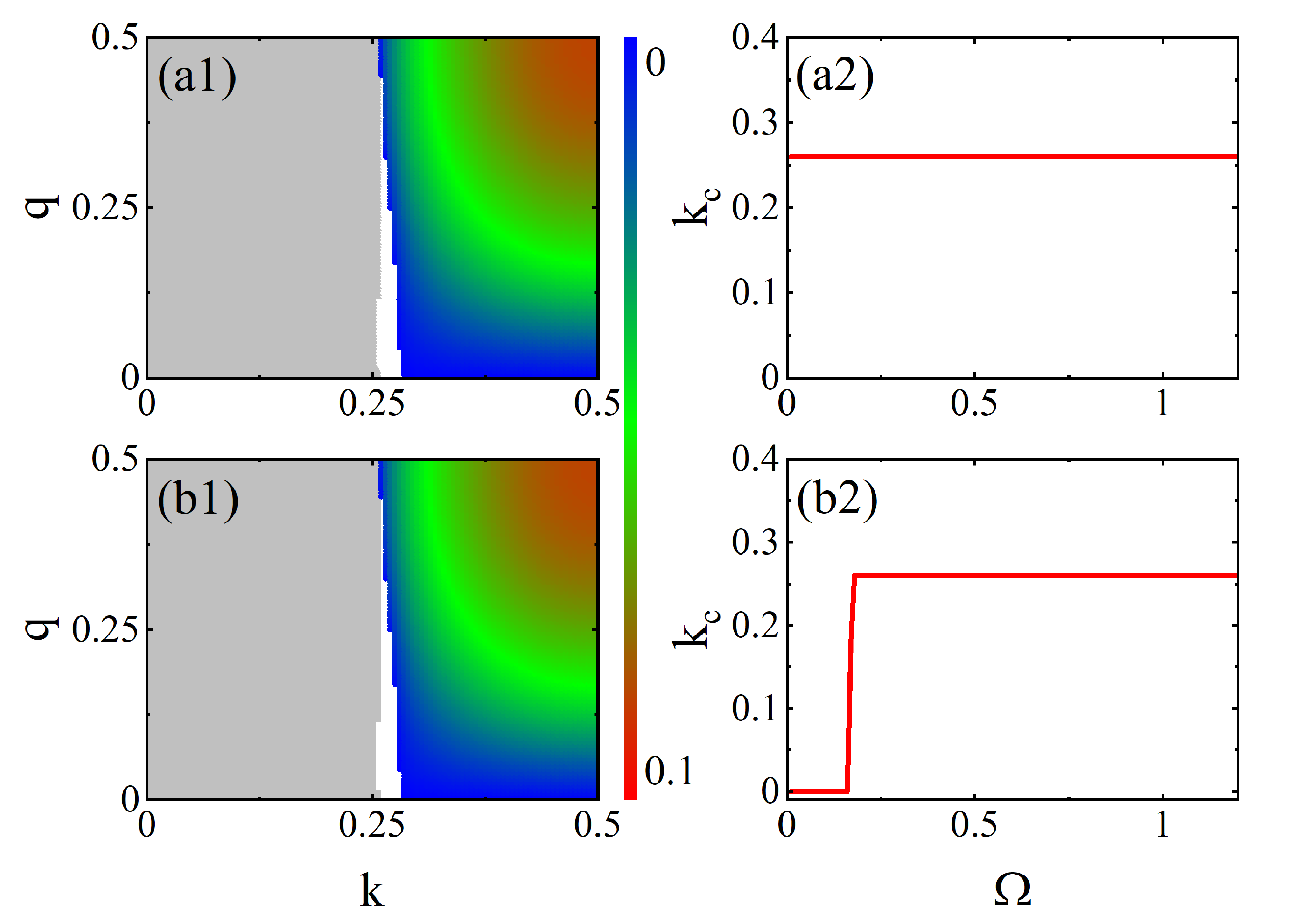}
	\caption{The instability of Bloch states in the Rabi-coupled spin-independent lattice,  $v_1=v_2=0.5$.  (a1)
		Unstable regimes in the ($q,k$) plane with $g=0.15,\gamma=1/3$.   The light gray area represents stable, out of the light gray area is Landau unstable and the dark colored shadow area indicates dynamical instability with the color scale labeling the amplitude of the growth rate which is defined as the maximum of the imaginary parts of $\omega$ in Eq.~(\ref{BdG}), $\Omega=1$.
	(a2) The critical quasimomenta $k_c^{D,L}$ as a function of $\Omega$ for $g=0.15,\gamma=1/3$. $k_c^{D}$ and $k_c^{L}$ overlap. (b1) and (b2) show same quantities as in (a1) and (a2) respectively, but in (b1) and (b2) the parameters are $g=0.05,\gamma=3$.
	}
	\label{Oneinstability}
\end{figure}
We analyze the instabilities of Bloch states in the presence of the Rabi coupling. The spin-independent lattice defines first Brillouin zone with edges at $k=\pm 1/2$. All interaction coefficients we consider are repulsive. Due to this, nonlinear Bloch spectrum $\mu(k)$ is displaced upwards with respect to the linear spectrum (at $g_{ij}=0$). A typical nonlinear Bloch spectrum is demonstrated in Fig.~\ref{Onespectrum}(a). Comparing with the linear spectrum, there is no obvious change of the structure in the nonlinear spectrum. Fig.~\ref{Onespectrum}(b) and (c) show the profiles of nonlinear Bloch states in the lowest band at Brillouin zone center and edge respectively. Two components stay in each well of $\cos(x)$ and spatially overlap. We study the stability of these states in the lowest band with quasimomentum $k \in (-1/2,1/2]$ for different $\Omega$. Two representative results are shown in Fig.~\ref{Oneinstability}(a1) and (b1) for $g=0.15,\gamma=1/3$ and $g=0.05,\gamma=3$ respectively with $\Omega=1$. For the Bloch state with a fixed $k$, the quasimomentum of perturbations $q$ can  choose the values of $ (-1/2,1/2]$. With the given values of $(q,k)$, we calculate equation~(\ref{BdG}) and the eigenvalue of $\tau_z  \mathcal{H}_\text{BdG}$ to judge the stability of the corresponding Bloch state.  Considering the symmetry $k\rightarrow -k$ and $q\rightarrow -q$, we only show the results with parameters $(q,k)\in [0,1/2]$. In the ($q,k$) plane,
in Fig.~\ref{Oneinstability}(a1) and (b1),  the stable states are represented by the light gray areas.  Out of the gray areas denote Landau unstable states and dark colored areas denote dynamically unstable modes. The color scale is used to label the growth rate of the dynamical instability, which is defined as the maximum of imaginary parts of $\omega$ in Eq.~(\ref{BdG}). The structure of the dynamical instability in the ($q,k$) plane is analogues to two components without the Rabi coupling~\cite{Jin2005,Ruostekoski2007}. From these two plots, we can know that there is a sharp boundary between stable and unstable Bloch states. From the boundary, we define the critical quasimomenta, $k_c^{D(L)}$. Beyond the critical values, $|k|>k_c^{D(L)} $, the Bloch states are dynamically unstable (Landau unstable). The dependence of $k_c^{D(L)}$ on the Rabi coupling is presented in Fig.~\ref{Oneinstability}(a2) and (b2). For the coefficients satisfying $\gamma<1$, the Rabi coupling has no effect on the critical quasimomenta, i.e., $k_c^{D(L)}$ are constant as a function of the Rabi coupling [see Fig.~\ref{Oneinstability}(a2)]. While, for the case of $\gamma>1$ shown in Fig.~\ref{Oneinstability}(b2), the Rabi coupling changes the stability of the Bloch states.  When $\Omega=0$, the critical quasimomenta $k_c^{D(L)}=0$, this result is consistent with the findings in Ref.~\cite{Jin2005,Liu2007,Ruostekoski2007}.  The phase separation instability due to $\gamma>1$ makes all Bloch states in the lowest bands unstable. A small $\Omega$ is useless and the critical quasimomenta remain at $0$. Once $\Omega$ is beyond a threshold value, $k_c^{D(L)}$ abruptly change into finite constants. Therefore, the Rabi coupling can overcome the phase separation instability and can stabilize the Bloch states around Brillouin zone center. Such stabilization is reminiscent of the modification of phase separation criterion by the Rabi coupling in homogeneous BECs.  Further increasing $\Omega$,  $k_c^{D(L)}$  always keep constant $\approx 0.25$ which is the same value as the case of $\gamma<1$ shown in Fig.~\ref{Oneinstability}(a2). The constant value approximately corresponds to the middle between Brillouin zone center and edge.

\begin{figure}[t]
	\includegraphics[width=0.45\textwidth]{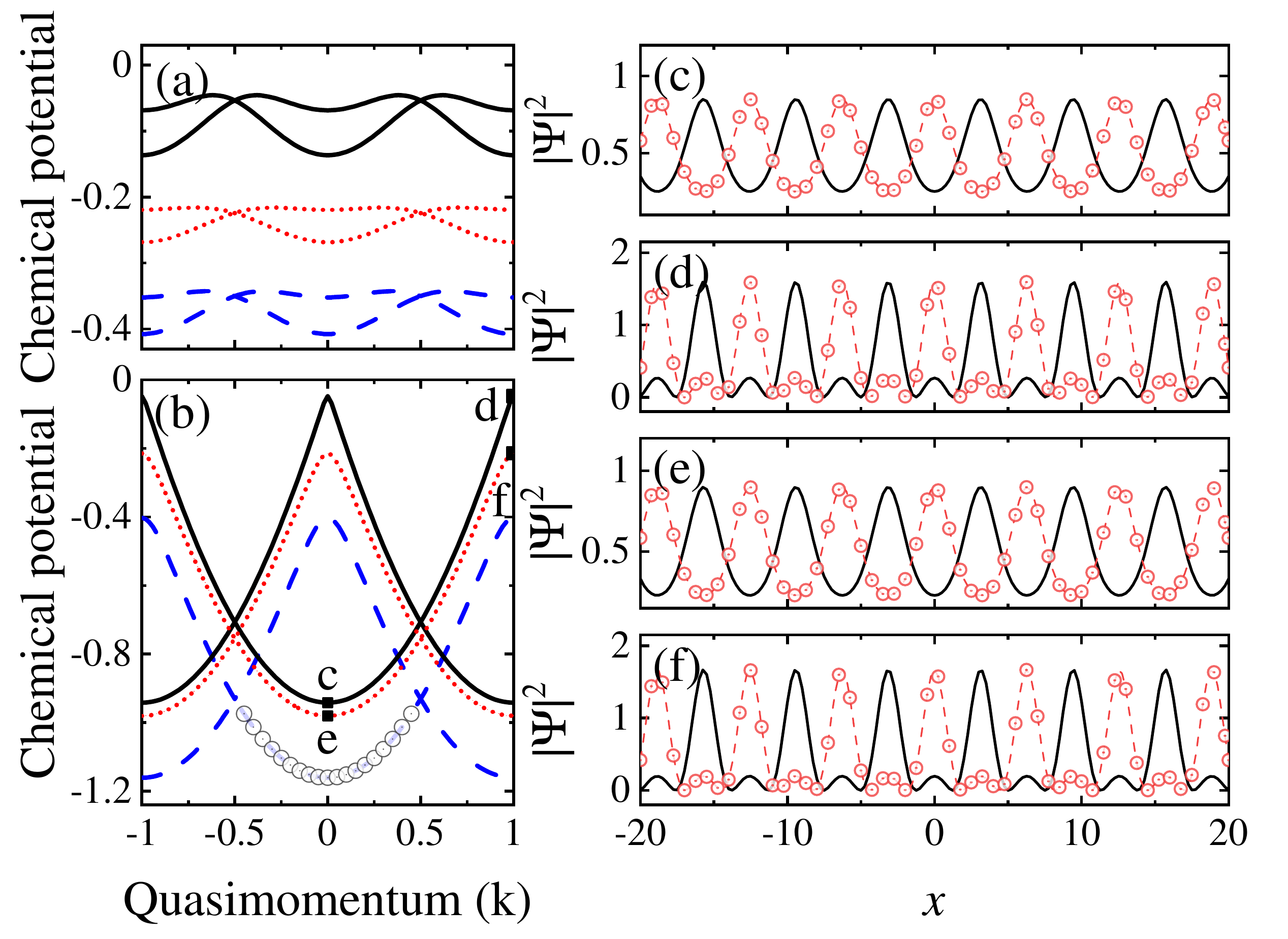}
	\caption{Nonlinear spectrum and nonlinear Bloch states in the Rabi-coupled Zeeman lattice, $v_1=-v_2=0.5$. (a)
		Linear (dashed blue lines) and nonlinear
		spectrum with $g=0.05,\gamma=3$ (dotted red lines) and with $g=0.15,\gamma=1/3$ (solid black line). Only the lowest two bands are shown,  $\Omega=0.05$. (b) demonstrates same quantities as in (a) with $\Omega=1$. Circles are the explicit dispersion relation $k^2/(2m^*)$ after a proper vertical displacement in the neighborhood of $k=0$ with effective mass $m^*=1$. The nonlinear Bloch states labeled by squares are illustrated in (c)-(f) where black lines are $|\Psi_1|^2$ and dotted red lines are  $|\Psi_2|^2$.}
	\label{Twospectrum}
\end{figure}


\section{ Zeeman lattice with $v_1=-v_2$}
\label{ZeemanLattice}

When $v_1=-v_2=v$, the lattice is a Zeeman lattice. The Zeeman lattice has been used to effectively manipulate interactions between components~\cite{Ostrovskaya2004}. The instability of Bloch states in the Zeeman lattice without the Rabi coupling have been studied in Ref.~\cite{Hooley2007}. It was found that for miscible interactions $\gamma<1$ the instability is qualitatively similar to the result in the single component BEC. For immiscible interactions $\gamma>1$ all Bloch states in the lowest band are unstable when the depth $v$ is small and a large depth can stabilize the Bloch states around Brillouin zone center~\cite{Hooley2007}.

  \begin{figure}[b]
  	\includegraphics[width=0.45\textwidth]{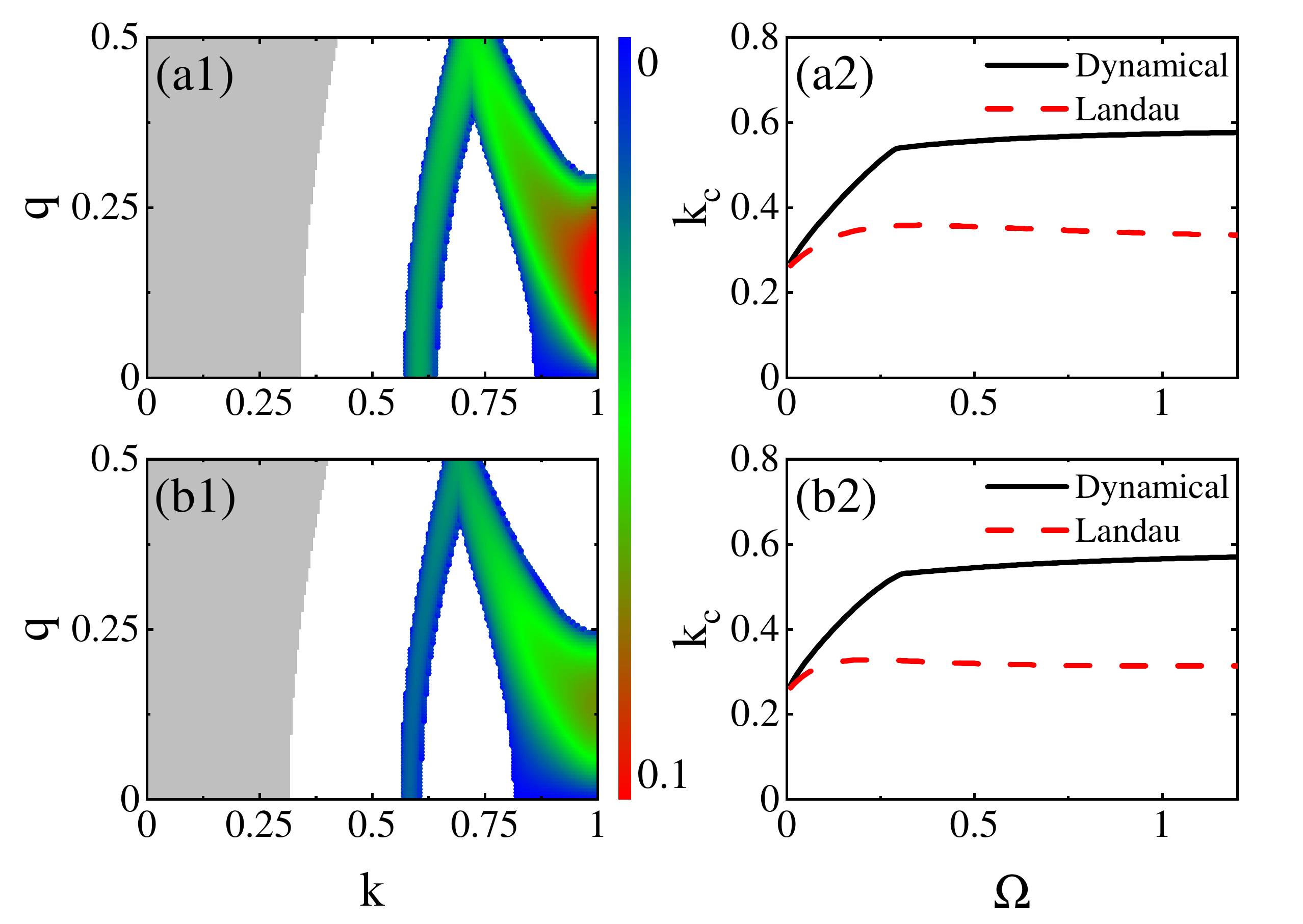}
  	\caption{The instability of Bloch states in the Rabi-coupled Zeeman lattice, $v_1=-v_2=0.5$. All panels show the same quantities as in Fig.~\ref{Oneinstability}.}
  	\label{Twoinstability}
  \end{figure}

We study the effect of the Rabi coupling in the Zeeman lattice. In the spinor basis of $\Psi=(\Psi_1, \Psi_2)^T$, the linear part of the Hamiltonian according to Eq.~(\ref{GP}) is
\begin{equation}
H_\text{lin}=-\frac{1}{2}\frac{\partial^2}{\partial x^2} +v\cos(x)\sigma_z+\frac{\Omega}{2} \sigma_x,
\end{equation}
where $\sigma$ are Pauli matrices. The non-commutation of Pauli matrices gives rise to a competition between the Zeeman lattice and the Rabi coupling in $H_\text{lin}$. The competition endows a half-period translational symmetry,
\begin{equation}
T=\sigma_xe^{i\pi \hat{p}},
\end{equation}
with $\hat{p}=-i\partial  /\partial x$ and $[T,H_\text{lin}]=0$~\cite{Wilkens1991,Larson2009}. This symmetry extends first Brillouin zone to $k \in (-1,1]$. With the symmetry,  $H_\text{lin}$ can be block-diagonalized into two uncoupled subsystems, so that the energy spectrum of  $H_\text{lin}$ can be grouped into two sets belonging to the eigenvalues of the subsystems~\cite{Larson2009}.
Fig.~\ref{Twospectrum}(a) and (b) demonstrate the lowest two  bands of  $H_\text{lin}$ (dashed blue lines in the bottom). One of them corresponds to the eigenvalues of one of the subsystems. Band crossings exist at $k=\pm 1/2$ due to the uncoupling of the subsystems. These two bands are same if displacing the quasimomentum by 1,  because of the standard translational symmetry $e^{i2\pi\hat{p}}$. From  Fig.~\ref{Twospectrum}(a) and (b), it is noticed that the Rabi coupling alters the geometries of the linear spectrum.   While, the interactions do not dramatically change the structure of the linear spectrum [see Fig.~\ref{Twospectrum}(a) and (b)]. With the interactions, the band crossings at $k=\pm 1/2$ and the extension of first Brillouin zone still exist. Nevertheless, the interactions bring interesting features to the instability of Bloch states. Taking into account the displacement symmetry between the lowest two bands, we only focus on the band having a minimum at $k=0$. The typical profiles of nonlinear Bloch states in the focused band labeled by squares in Fig.~\ref{Twospectrum}(b) are presented in Fig.~\ref{Twospectrum}(c)-(f). These Bloch states fully reflect the symmetry $T$. The first component is centered at the minima of $\cos(x)$, while the second component is accumulated at the minima of $-\cos(x)$. The spatial separation between the two components at $k=0$ is very obvious [see Fig.~\ref{Twospectrum}(c) and (e)]. While at $k=1$, there is a small overlapped part in Fig.~\ref{Twospectrum}(d) and (f). Even though the first Brillouin zone is extended, the Bloch states at $k=1$ still have the period of $2\pi$.

 \begin{figure}[b]
 	\includegraphics[width=0.45\textwidth]{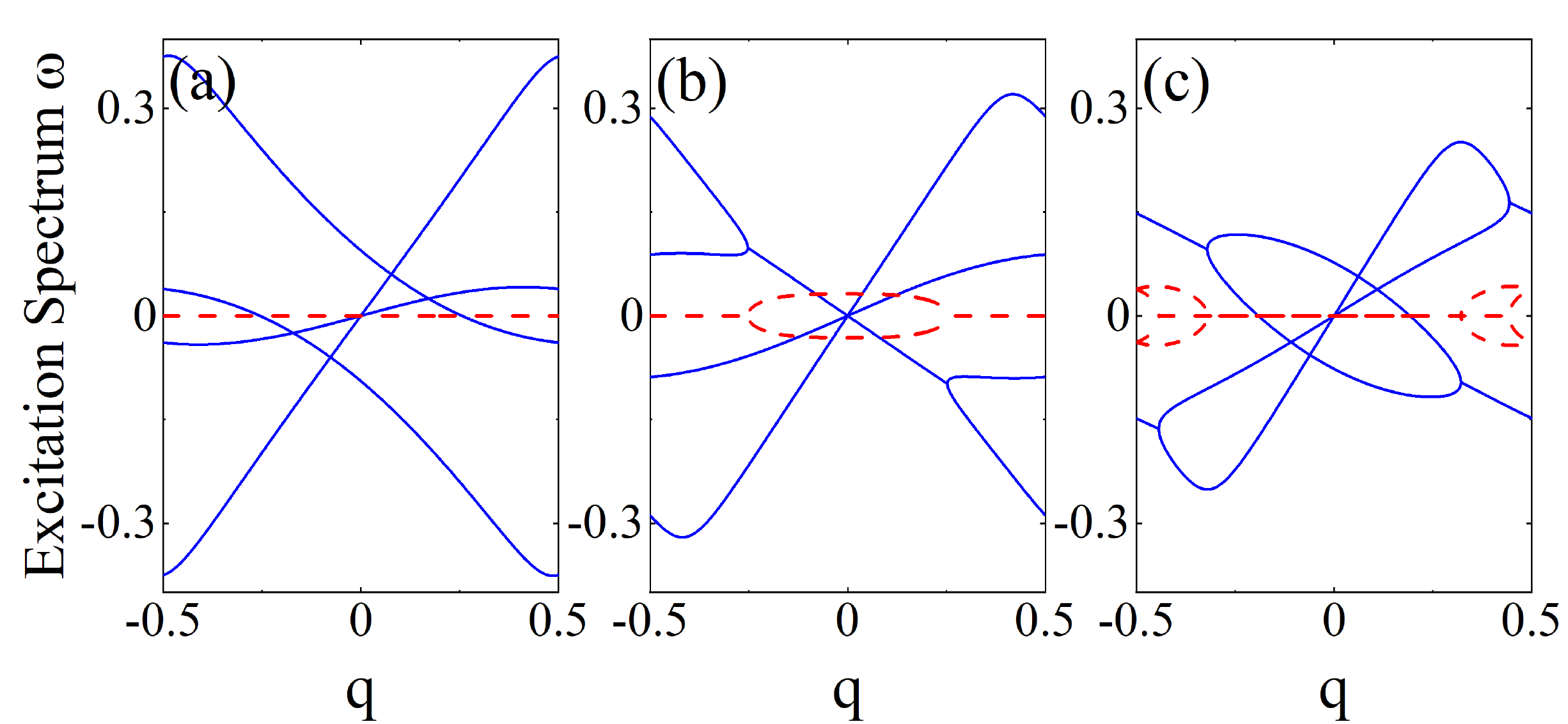}
 	\caption{The collective excitation spectrum $\omega(q)$ of Bloch states at $k=0.5$ (a), 0.6 (b) and 0.7 (c) in the Rabi-coupled Zeeman lattice. The solid blue lines are the real part of $\omega$ and dashed red lines are the imaginary part. }
 	\label{Twoexcitation}
 \end{figure}

The instability results are presented in
 Fig.~\ref{Twoinstability}. In the presence of the Rabi coupling, the dynamical instability features a $\Lambda$ like shape in the $(q,k)$ plane [see the colored areas in Fig.~\ref{Twoinstability}(a1) and (b1)]. The right-hand branch of the $\Lambda$ is similar with the structure without the Rabi coupling as found in~\cite{Hooley2007} and the branch on left-hand side is new due to the Rabi coupling. The development of dynamically unstable modes in collective excitations is shown in Fig.~\ref{Twoexcitation}. The collective excitation spectrum has the symmetry, $\omega(q)=-\omega(-q)$. The nonlinear Bloch state at $k=0.5$ is dynamically stable, and all spectrum is real-valued [see Fig.~\ref{Twoexcitation}(a)]. For the state at $k=0.6$, the imaginary modes appear around $q=0$, which is demonstrated in Fig.~\ref{Twoexcitation}(b), this instability corresponds to the left-hand branch of the $\Lambda$ shown in Fig.~\ref{Twoinstability}(a1). These imaginary modes move from the $q=0$ towards to the edges $q=\pm 1/2$ with increasing $k$. At $k=0.7$ in Fig.~\ref{Twoexcitation}(c), more imaginary modes arise and they correspond to the right-hand branch of the $\Lambda$. From the results in Fig.~\ref{Twoinstability}(a1) an d(b1), it can be seen that the separation of dynamical (colored areas) and Landau (out of the gray areas) is very obvious. Such separation is more evident for the critical quasimomenta $k_c^{D,L}$ shown in Fig.~\ref{Twoinstability}(a2) and (b2). In the Rabi-coupled spin-independent lattice in the previous section, $k_c^{D}$ and  $k_c^{L}$ coincide, while in the Zeeman lattice both $k_c^{D,L}$ increase as a function of $\Omega$ when $\Omega$ is small.  When $\Omega$ is large, $k_c^{D,L}$ saturate to different values.  When $\Omega=0$, although the half-period translational symmetry $T$ still commutes with the linear Hamiltonian, it is trivial due to the uncoupling between two components. The first Brillouin zone edges are not extended and are at $k=\pm 1/2$. $k_c^D$ is nearby $1/4$ laying at the middle between the center and edge of Brillouin zone, which is consistent with the result in  Ref.~\cite{Hooley2007}. In the presence of the Rabi coupling, the first Brillouin zone edges are extended at $k=\pm 1$. We find that $k_c^D$ follows the extension and the saturated value is nearby the middle of the extended edge, $k_c^D \approx 0.5$.


Unlike the dynamical instability,  the Landau instability does not follow the extension of Brillouin zone, $k_c^L$ is always around $0.3$, which is demonstrated in Fig.~\ref{Twoinstability}(a2) and (b2). We provide a physical insight into the understanding of this behavior.  In the long-wavelength limit, i.e., $k\rightarrow 0$ and $q\rightarrow 0$, the spatial modulation of the Zeeman lattice can be approximately neglected and its other effects can be absorbed into the effective mass $m^*$~\cite{Jin2005}.  When $\Omega$ is small, the effective mass is much larger than the original mass (which is 1 in dimensionless unit), $m^*>1$; if $\Omega$ takes a large value, $m^*=1$, this can be seen from the fitting of the spectrum around $k=0$ in Fig.~\ref{Twospectrum}(b). In the long-wavelength limit, the system can be approximately described by the two-coupled spatially homogeneous BEC with dressed mass $m^*$. For a two-coupled homogeneous BEC, it has been analytically uncovered that the lowest collective excitation branch is independent of $\Omega$ and the Landau critical velocity is $v_L=\sqrt{(g+g_{12})n/(m^*)}$ where $n$ is the atomic density~\cite{Abad2013,Momme2020}.
We approximate the density by an average density $n=0.5$ from the nonlinear Bloch states shown in Fig.~\ref{Twospectrum}(c) and (e).
Applying $m^*=1$ and $n=0.5$, we know $v_L= 0.32$ using interaction parameters of Fig.~\ref{Twoinstability}. According to the Landau criterion of superfluidity, the critical value is $k_c^L=v_L$~\cite{Danshita2007}, and it is independent of $\Omega$ when $\Omega$ takes dominating values.

The large separation between  $k_c^L$ and $k_c^D$ in Figs.~\ref{Twoinstability}(a2) and (b2) is due to the extension of Brillouin zone  originating from the half-period translational symmetry. The separation between $k_c^D$ and $k_c^L$ can even exist in a single-component BEC if the lattice depth is small~\cite{Wu2001,Modugno2004}, and if the depth is large, two values are coincide~\cite{Smerzi2002}. However,  the separation is not obvious in all existed systems studied in literature. Above significant separation is a unique feature of the Rabi-coupled Zeeman lattice which provides an ideal platform to study dynamical and Landau instabilities of lattice superfluids separately.

We also observe that the difference between the cases with $\gamma=1/3$ in Fig.~\ref{Twoinstability}(a1), (a2) and $\gamma=3$ in Fig.~\ref{Twoinstability}(b1), (b2) is not obvious. This is because we use a large Zeeman lattice depth ($v=0.5$). The competition between the Zeeman lattice and the Rabi coupling completely dominates over the effect of interactions.

\begin{figure}[t]
	\includegraphics[width=0.45\textwidth]{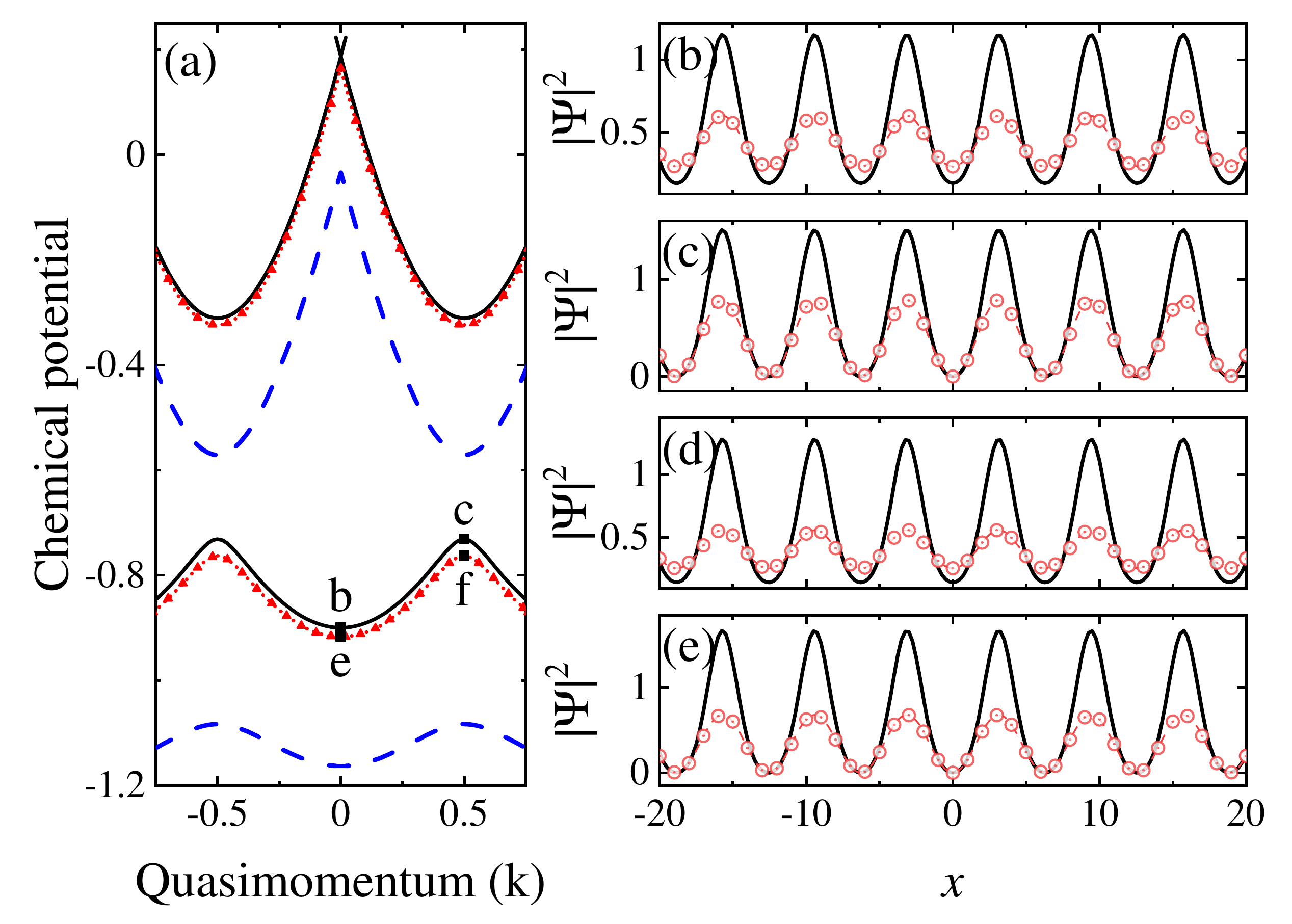}
	\caption{Nonlinear spectrum and nonlinear Bloch states in the Rabi-coupled matter grating. $v_1=0.5$, $v_2=0$ and $\Omega=1$. All panels show the same quantities as in Fig.~\ref{Twospectrum}(b)-(f).
	}
	\label{Threespectrum}
\end{figure}

\section{Matter grating with $v_1 \ne 0$, $v_2=0$}
\label{Mattergrating}

If only the one component feels optical lattices (i.e., $v_1 \ne 0$ and $v_2=0$), its density is distributed periodically due to Bloch states. Via the inter-component interactions, the other component suffers a matter grating.  The instability of Bloch states in the matter grating without the Rabi coupling has been thoroughly identified in Ref.~\cite{Barontini2009}. Their two components are from different species of atoms, so the masses of  two components are not equal. They found that the phase separation instability for $\gamma>1$ still makes all Bloch states unstable and furthermore the dynamical instability has complex structures in the $(q,k)$ plane~\cite{Barontini2009}.

We incorporate
the Rabi coupling into the matter grating. The nonlinear spectrum shown in Fig.~\ref{Threespectrum}(a) is very different from that in the Zeeman lattice (in Fig.~\ref{Twospectrum}) even though both lattices are spin-dependent. There is no half-period translational symmetry in the Rabi-coupled matter grating, so the first Brillouin zone is $k\in (-1/2,1/2]$. In fact, the lowest band is similar to that of the spin-independent lattice demonstrated in Fig.~\ref{Onespectrum}(a). However, the density profiles of Bloch states are different from the spin-independent case. Two components are still centered at the minima of $\cos(x)$, but have different amplitudes. The first component is always larger than the second component [see Fig.~\ref{Threespectrum}(b)-(e)].

\begin{figure}[t]
	\includegraphics[width=0.45\textwidth]{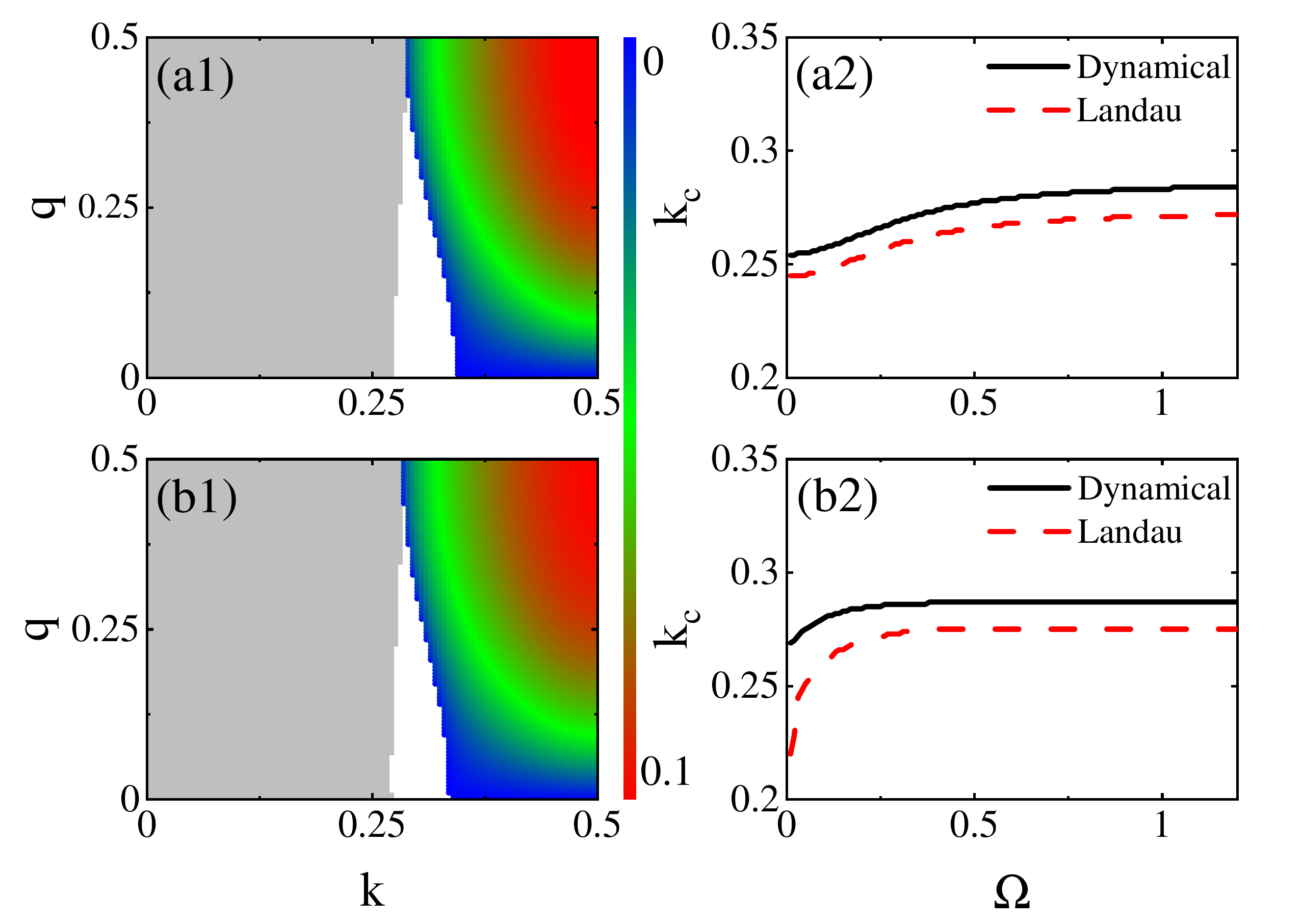}
	\caption{The instability of Bloch states in the Rabi-coupled matter grating, $v_1=0.5$ and $v_2=0$. All panels show the same quantities as in Fig.~\ref{Oneinstability} and Fig.~\ref{Twoinstability}.
	}
	\label{Threeinstability}
\end{figure}

The instability results for $\Omega=1$ are described in the ($q,k$) plane  in Fig.~\ref{Threeinstability}(a1) and (b1) for $\gamma=1/3$ and $\gamma=3$ respectively. The structures of the instabilities are similar to these of the spin-independent lattice. However,  the critical quasimomenta of dynamical and Laudau instabilities are not equal.  $k_c^D$ is always slightly larger than $k_c^L$, which is presented in  Fig.~\ref{Threeinstability}(a2) and (b2). $k_c^{D,L}$ increase as a function of $\Omega$ and then slowly saturate for $\gamma=1/3$, while for $\gamma=3$ they rapidly saturate. In the spinor basis, the lattice corresponds to $v_1/2 \cos(x)+v_1/2\cos(x) \sigma_z$ including a spin-independent part and a Zeeman lattice part. It looks that the behaviors of $k_c^{D,L}$ neutralize the features of the spin-independent lattice and the Zeeman lattice. On the other hand, the energy functional of interactions and the Rabi coupling is $ g/2 (1+\gamma) ( |\Psi_1|^2+|\Psi_2|^2 )^2 + g/2 (1-\gamma) \langle \sigma_z \rangle^2 + \Omega \langle \sigma_x \rangle $, where $\langle \sigma_z \rangle =  |\Psi_1|^2|-|\Psi_2|^2 $ and $\langle \sigma_x \rangle =  \Psi_1\Psi_2^*+\Psi_1^*\Psi_2 $. When $\gamma>1$, the minimization of the second term in the energy functional needs to $\langle \sigma_z \rangle  \ne 0$. Oppositely, the minimization of the Rabi coupling prefers to $\langle \sigma_x \rangle \ne 0$ and $\langle \sigma_z \rangle =0$. However, when $\gamma<1$, the minimization of the second term requires $\langle \sigma_z \rangle =0$, having the same purpose as the Rabi coupling.  Therefore, the interactions with $\gamma>1$ and the Rabi coupling constitute a direct competition. The effect of the Rabi coupling may be more distinct for $\gamma>1$ and gives rise to the dramatical dependence of $k_c^{D,L}$ on $\Omega$ in Fig.~\ref{Threeinstability}(b2).

\section{conclusion}
\label{Conclusion}

In conclusion, we systematically demarcate  the instability regimes of lattice superfluids (represented by Bloch states) for a two-coupled BEC. Without the Rabi coupling,  the phase separation instability makes superfluids unstable both dynamically and energetically. We find that the Rabi coupling can stabilize superfluids around Brillouin zone center. Such stabilization is very important for experimental realizations. In experiments, BECs are adiabatically loaded into optical lattices. The experimental implementation prepares BECs into the lowest energy state which is at Brillouin zone center. If the lowest energy state is unstable, adiabatic process will destroy loading. The Rabi coupling is a possible way to make loading successful.

Both dynamical and Landau instabilities may exist in a same BEC Bloch state~\cite{Wu2001,Smerzi2002,Menotti2003,Modugno2004}.  A Bloch state that is dynamically unstable must be Landau unstable. However, Landau unstable Bloch state does not necessarily mean dynamically unstable. We uncover that there exists a significant regime where Bloch states are Landau unstable but dynamically stable. Such unique feature is provided by the Rabi-coupled Zeeman lattice. The extension of first Brillouin zone to double its value is due to a specific half-period translational symmetry. The regime of dynamical instability follows the extension. While,  Landau instability conservatively happens around Brillouin zone center. Therefore, the separation between Landau and dynamical instabilities becomes very significant. The Rabi-coupled Zeeman lattice represents an outstanding system to identify the structures of Landau instability which can be separated from dynamical instability.

\begin{acknowledgments}
	
This work is supported by National Natural Science Foundation of China with Grants Nos.~11974235 and 11774219.

\end{acknowledgments}


\begin{thebibliography}{100}

\bibitem{Morsch2006}
O.~Morsch and M.~Oberthaler, Dynamics of Bose-Einstein condensates in optical lattices,
Rev.~Mod.~Phys.~{\bf 78}, 179 (2006).

\bibitem{Bloch2008}
I.~Bloch, J.~Dalibard, and W.~Zwerger, Many-body physics with ultracold gases,  Rev.~Mod.~Phys.~{\bf 80}, 885 (2008).

\bibitem{Georgescu2014}
I.~M.~Georgescu, S.~Ashhab, and F.~Nori, Quantum simulation, Rev.~Mod.~Phys.~{\bf 86}, 153 (2014).

\bibitem{Wu2001}
B.~Wu and Q.~Niu, Landau and dynamical instabilities of the superflow of Bose-Einstein condensates in optical lattices,
Phys.~Rev.~A~{\bf 64}, 061603(R) (2001).

\bibitem{Burger2001}
S.~Burger, F.~S.~Cataliotti, C.~Fort, F.~Minardi, M.~Inguscio, M.~L.~Chiofalo, and M.~P.~Tosi, Superfluid and dissipative dynamics of a Bose-Einstein condensate in a periodic optical potential,  Phys.~Rev.~Lett.~{\bf 86}, 4447 (2001).

\bibitem{Smerzi2002}
A.~Smerzi, A.~Trombettoni, P.~G.~Kevrekidis, and A.~R.~Bishop, Dynamical superfluid-insulator transition in a chain of weakly coupled Bose-Einstein condensates,  Phys.~Rev.~Lett.~{\bf 89}, 170402 (2002).

\bibitem{Konotop2002}
V.~V.~Konotop and M.~Salerno, Modulational instability in Bose-Einstein condensates in optical lattices,
Phys.~Rev.~A {\bf 65}, 021602(R) (2002).

\bibitem{Machholm2003}
M.~Machholm, C.~J.~Pethick, and H.~Smith, Band structure, elementary excitations, and stability of a Bose-Einstein condensate in a periodic potential, Phys.~Rev.~A~{\bf 67} 053613 (2003).

\bibitem{Wu2003}
B.~Wu and Q.~Niu, Superfluidity of Bose-Einstein condensate in an optical lattice: Landau-Zener tunnelling and dynamical instability,
New~J.~Phys.~{\bf 5}, 104 (2003).

\bibitem{Menotti2003}
C.~Menotti, A.~Smerzi, and A.~Trombettoni, Superfluid dynamics of a Bose-Einstein condensate in a periodic potential,
New~J.~Phys.~{\bf 5}, 112 (2003).

\bibitem{Fallani2004}
L.~Fallani, L.~De Sarlo, J.~E.~Lye, M.~Modugno, R.~Saers, C.~Fort, and M.~Inguscio, Observation of dynamical instability for a Bose-Einstein condensate in a moving 1D optical lattice,
Phys.~Rev.~Lett.~{\bf 93}, 140406 (2004).

\bibitem{Modugno2004}
M.~Modugno, C.~Tozzo, and F.~Dalfovo, Role of transverse excitations in the instability of Bose-Einstein condensates moving in optical lattices,  Phys.~Rev.~A~{\bf 70} 043625 (2004).


\bibitem{Sarlo2005}
L.~De Sarlo, L~ Fallani, J.~E.~Lye, M.~Modugno, R.~Saers, C.~Fort, and M.~Inguscio, Unstable regimes for a Bose-Einstein condensate in an optical lattice,  Phys.~Rev.~A~{\bf 72}, 013603 (2005).

\bibitem{Danshita2007}
I.~Danshita and S.~Tsuchiya, Stability of Bose-Einstein condensates in a Kronig-Penney potential,
Phys.~Rev.~A~{\bf 75}, 033612 (2007).

\bibitem{Hui2011}
H.-Y.~Hui, R.~Barnett, R.~Sensarma, and S.~D.~Sarma, Instabilities of bosonic spin currents in optical lattices
Phys.~Rev.~A~{\bf 84} 043615 (2011).

\bibitem{Chen2011}
Z.~Chen and B.~Wu,  Bose-Einstein condensate in a honeycomb optical lattice: Fingerprint of superfluidity at the Dirac point,
Phys.~Rev.~Lett.~{\bf 107}, 065301 (2011).

\bibitem{Shaoliang2013}
S.-L.~Zhang, Z.-W.~ Zhou, and B. Wu, Superfluidity and stability of a Bose-Einstein condensate with periodically modulated interatomic interaction, Phys.~Rev.~A~{\bf 87}, 013633 (2013).

\bibitem{Xu2013}
Y.~Xu, Z.~Chen, H.~Xiong, W.~V.~Liu, and B.~Wu, Stability of p-orbital Bose-Einstein condensates in optical checkerboard and square lattices, Phys.~Rev.~A~{\bf 87}, 013635 (2013).

\bibitem{Dasgupta2016}
R.~Dasgupta, B.~P.~Venkatesh, and G.~Watanabe, Attraction-induced dynamical stability of a Bose-Einstein condensate in a nonlinear lattice, Phys.~Rev.~A~{\bf  93}, 063618 (2016).


\bibitem{Chen2010}
Z.~Chen and B.~Wu, Stability of Bose-Einstein condensates in two-dimensional optical lattices,
Phys.~Rev.~A~{\bf  81}, 043611 (2010).

\bibitem{Denschlag2002}
J.~Hecker Denschlag, J.~ E.~Simsarian, H.~ H\"affner, C.~McKenzie, A.~Browaeys, D.~Cho, K.~Helmerson, S.~ L.~Rolston, and W.~ D.~Phillips, A Bose-Einstein condensate in an optical lattice, J.~Phys.~B:~At.~Mol.~Opt.~Phys.~{\bf 35}, 3095 (2002).




\bibitem{Ozeri2005}
R.~Ozeri, N.~Katz, J.~Steinhauer, and N.~Davidson, Bulk Bogoliubov excitations in a Bose-Einstein condensate,  Rev.~Mod.~Phys.~{\bf 77}, 187 (2005).

\bibitem{Fabbri2009}
N.~Fabbri, D.~Cl\'ement, L.~Fallani, C.~Fort, M.~Modugno, K.~M.~R.~van der Stam, and M.~Inguscio, Excitations of Bose-Einstein condensates in a one-dimensional periodic potential,
Phys.~Rev.~{\bf A} 79, 043623 (2009).







\bibitem{Jin2005}
G.-R.~Jin, C.~K.~Kim, and K.~Nahm, Modulational instability of two-component Bose-Einstein condensates in an optical lattice,
Phys.~Rev.~A~{\bf 72}, 045601 (2005).

\bibitem{Hooley2007}
S.~Hooley and K.~A.~Benedict, Dynamical instabilities in a two-component Bose-Einstein condensate in a one-dimensional optical lattice,  Phys.~Rev.~A~{\bf 75}, 033621 (2007).

\bibitem{Liu2007}
X.~P.~Liu, Excitation spectrum and phase separation of double Bose-Einstein condensates in optical lattices,
Phys.~Rev.~A~{\bf 76}, 053615 (2007).


\bibitem{Ruostekoski2007}
J.~Ruostekoski and Z.~Dutton, Dynamical and energetic instabilities in multicomponent Bose-Einstein condensates in optical lattices,  Phys.~Rev.~A~{\bf 76}, 063607 (2007).

\bibitem{Barontini2009}
G.~Barontini and M.~Modugno, Instabilities of a matter wave in a matter grating,  Phys.~Rev.~A~{\bf 80}, 063613 (2009).

\bibitem{Huang2010}
J.-S.~ Huang, Z.-W.~Xie, M. Zhang, and L.-F.~Wei, Modulational instability of two-component Bose-Einstein condensates in
an optical lattice formed inside a cavity, J.~Phys.~B:~At.~Mol.~Opt.~Phys.~{\bf 43}, 065305 (2010).

\bibitem{Watanabe2018}
G.~Watanabe and Y.~Zhang, Stabilization of nonlinear lattices: A route to superfluidity and hysteresis,
Phys.~Rev.~A~{\bf 98}, 013625 (2018).


\bibitem{Wright2009}
K.~C.~Wright, L.~S.~Leslie, A.~Hansen, and N.~P.~Bigelow, Sculpting the vortex state of a spinor BEC,  Phys.~Rev.~Lett.~{\bf 102}, 030405 (2009).

\bibitem{Hamner2013}
C.~Hamner, Y.~Zhang, J.~J.~Chang, C.~Zhang, and P.~Engels,  Phase winding a two-component Bose-Einstein condensate in an elongated trap: experimental observation of moving magnetic orders and dark-bright solitons,
Phys.~Rev.~Lett.~{\bf 111}, 264101(2013).

\bibitem{Lundblad2008}
N.~Lundblad, P.~J.~Lee, I.~B.~Spielman, B.~L.~Brown, W.~D.~Phillips, and J.~V.~Porto, Atoms in a radio-frequency-dressed optical lattice,
Phys.~Rev.~Lett.~{\bf 100}, 150401 (2008).


\bibitem{Matthew1998}
M.~R.~Matthews, D.~S.~Hall, D.~S.~Jin, J.~R.~Ensher, C.~E.~Wieman, E.~A.~Cornell, F.~Dalfovo, C.~Minniti, and S.~Stringari, Dynamical response of a Bose-Einstein condensate to a discontinuous change in internal state,
Phys.~Rev.~Lett.~{\bf 81}, 243 (1998).









\bibitem{Lee2009}
C.~Lee, Universality and anomalous mean-field breakdown of symmetry-breaking transitions in a coupled two-component Bose-Einstein condensate, Phys.~Rev.~Lett.~{\bf 102}, 070401 (2009).

\bibitem{Adhikari2009}
S.~K.~Adhikari and B.~A.~Malomed, Two-component gap solitons with linear interconversion, Phys.~Rev.~A~{\bf 79}, 015602 (2009).

\bibitem{Sabbatini2011}
J.~Sabbatini, W.~H.~Zurek, and M.~J.~Davis,Phase separation and pattern formation in a binary Bose-Einstein condensate, Phys.~Rev.~Lett.~{\bf 107}, 230402 (2011).

\bibitem{Zhan2014}
F.~Zhan, J.~Sabbatini, M.~J.~Davis, and I.~P.~McCulloch, Miscible-immiscible quantum phase transition in coupled two-component Bose-Einstein condensates in one-dimensional optical lattices,
Phys.~Rev.~A~{\bf 90}, 023630 (2014).

\bibitem{Usui2015}
 A.~Usui and H.~Takeuchi, Rabi-coupled countersuperflow in binary Bose-Einstein condensates, Phys.~Rev.~A~{\bf 91}, 063635 (2015).

\bibitem{Sartori2015}
A.~Sartori, J.~Marino, S.~Stringari, and A.~Recati, Spin-dipole oscillation and relaxation of coherently coupled Bose–Einstein condensates,
New~J.~Phys.~{\bf 17}, 093036 (2015).


\bibitem{Congy2016}
T.~Congy, A.~M.~Kamchatnov, and N.~Pavloff, Nonlinear waves in coherently coupled Bose-Einstein condensates
Phys.~Rev.~A~{\bf 93}, 043613 (2016).

\bibitem{Abad2013}
M.~Abad and A.~Recati, A study of coherently coupled two-component Bose-Einstein condensates, Eur.~Phys.~J.~D~{\bf 67}, 148 (2013).


\bibitem{Qu2017}
C.~Qu, M.~Tylutki, S.~Stringari, and L.~P. Pitaevskii, Magnetic solitons in Rabi-coupled Bose-Einstein condensates, Phys.~Rev.~A~{\bf 95}, 033614 (2017).

\bibitem{Bornheimer2017}
U.~Bornheimer, I.~Vasi\'c, and W.~Hofstetter, Phase transitions of the coherently coupled two-component Bose gas in a square optical lattice, Phys.~Rev.~A~{\bf 96},063623 (2017).

\bibitem{Uranga2018}
B.~Mencia~Uranga and A.~Lamacraft,  Infinite lattices of vortex molecules in Rabi-coupled condensates, Phys.~Rev.~A~{\bf 97}, 043609 (2018).

\bibitem{Baals2018}
C.~Baals, H.~Ott, J.~Brand, and A.~M.~Mateo, Nonlinear standing waves in an array of coherently coupled Bose-Einstein condensates, Phys.~Rev.~A~{\bf 98}, 053603 (2018).

\bibitem{Ihara2019}
K.~Ihara and K.~Kasamatsu,
Transverse instability and disintegration of a domain wall of a relative phase in coherently coupled two-component Bose-Einstein condensates, Phys.~Rev.~A~{\bf 100}, 013630 (2019).




\bibitem{Eto2020}
M.~Eto, K.~Ikeno, and M.~Nitta, Collision dynamics and reactions of fractional vortex molecules in coherently coupled Bose-Einstein condensates
Phys.~Rev.~Research~{\bf 2}, 033373 (2020).

\bibitem{Farolfi2020}
A.~Farolfi, D.~Trypogeorgos, C.~Mordini, G.~Lamporesi, and G.~Ferrari, Observation of magnetic solitons in two-component Bose-Einstein condensates, Phys.~Rev.~Lett.~{\bf 125}, 030401 (2020).

\bibitem{Fan2020}
Z.~Fan, Z.~Chen, Y.~Li, and B.~A.~Malomed, Gap and embedded solitons in microwave-coupled binary condensates,
Phys.~Rev.~A~{\bf  101}, 013607 (2020).



\bibitem{Momme2020}
M.~R.~Momme, O.~O.~Prikhodko, and Y.~M.~Bidasyuk, Dispersion relations and self-localization of quasiparticles in coupled elongated Bose-Einstein condensates,
Phys.~Rev.~A~{\bf 102}, 043316 (2020).





\bibitem{Lin2011}
Y.-J.~Lin, K.~Jim\'enez-Garc\'ia, and I.~B.~Spielman, Spin-orbit-coupled Bose-Einstein condensates, Nature, {\bf 471}, 83 (2011).

\bibitem{Hamner2015}
C.~Hamner, Y.~Zhang, M.~A.~Khamehchi, M.~J.~Davis, and P.~Engels, Spin-orbit-coupled Bose-Einstein condensates in a one-dimensional optical lattice,  Phys.~Rev.~Lett.~{\bf 114}, 070401 (2015).

\bibitem{Zhang2013}
Y.~Zhang and C.~Zhang, Bose-Einstein condensates in spin-orbit-coupled optical lattices: Flat bands and superfluidity,
Phys.~Rev.~A~{\bf 87}, 023611 (2013).

\bibitem{Toniolo2014}
D.~Toniolo and J.~Linder,  Superfluidity breakdown and multiple roton gaps in spin-orbit-coupled Bose-Einstein condensates in an optical lattice,  Phys.~Rev.~A~{\bf 89}, 061605(R) (2014).


\bibitem{Mandel2003}
O.~Mandel, M.~Greiner, A.~Widera, T.~Rom, T.~W.~H\"ansch, and I.~Bloch, Coherent transport of neutral atoms in spin-dependent optical lattice potentials,  Phys.~Rev.~Lett.~{\bf 91}, 010407 (2003).

\bibitem{Yang2017}
B.~Yang, H.-N.~Dai, H.~Sun, A.~Reingruber, Z.-S.~Yuan, and J.-W.~Pan, Spin-dependent optical superlattice, Phys.~Rev.~A~{\bf 96}, 011602(R) (2017).


\bibitem{Ostrovskaya2004}
E.~A.~Ostrovskaya and Y.~S.~Kivshar,
Localization of two-component Bose-Einstein condensates in optical lattices,  Phys.~Rev.~Lett.~{\bf 92}, 180405 (2004).

\bibitem{Wilkens1991}
M.~Wilkens, E.~Schumacher, and P.~Meystre, Band theory of a common model of atom optics, Phys.~Rev.~A~{\bf 44}, 3130 (1991).

\bibitem{Larson2009}
J.~Larson and J.-P.~ Martikainen, Coupled two-component atomic gas in an optical lattice,  Phys.~Rev.~A~{\bf 78}, 063618 (2008).




\end{thebibliography}

\end{document}